\documentclass[aps,prb,twocolumn,showpacs]{revtex4-1}

\usepackage{graphicx}

\begin{document}

\title{Dielectric, piezoelectric, and elastic properties of~BaTiO$_3$/SrTiO$_3$
ferroelectric superlattices from first principles}

\author{Alexander I. Lebedev}
\email[]{swan@scon155.phys.msu.ru}
\affiliation{Physics Department, Moscow State University, \\
119991 Moscow, Russia}

\date{\today}

\begin{abstract}
The effect of epitaxial strain on the phonon spectra, crystal structure,
spontaneous polarization, dielectric, piezoelectric, and elastic properties
of (001)-oriented ferroelectric (BaTiO$_3$)$_m$/(SrTiO$_3$)$_n$
superlattices ($m = n = {}$1--4) was studied using the first-principles
density-functional theory. The ground state of free-standing superlattices
is the monoclinic $Cm$ polar phase. Under the in-plane biaxial compressive
strain, it transforms to tetragonal $P4mm$ polar phase, and under the in-plane
biaxial tensile strain, it transforms to orthorhombic $Amm2$ polar phase.
When changing the in-plane lattice parameter, a softening of several optical
and acoustic modes appears at the boundaries between the polar phases,
and corresponding components of dielectric, piezoelectric, and elastic tensors
diverge critically. The comparison of the mixing enthalpy of disordered
Ba$_{0.5}$Sr$_{0.5}$TiO$_3$ solid solution modeled using two special
quasirandom structures SQS-4 with the mixing enthalpy of the superlattices
reveals a tendency of the BaTiO$_3$--SrTiO$_3$ system to short-range ordering
and shows that these superlattices are thermodynamically quite stable.
\end{abstract}


\maketitle

\section{Introduction}

The success in creating of ferroelectric superlattices with a layer thickness
controlled with an accuracy of one monolayer offers new opportunities for
design of new ferroelectric multifunctional materials with high spontaneous
polarization, Curie temperature, dielectric constant, and large dielectric
and optical nonlinearities. Because of many problems encountered in the growth
and experimental studies of ferroelectric superlattices, first-principles
calculations of their physical properties can be used to reveal new promising
fields of investigations and applications of these materials.

Earlier studies of thin epitaxial films of ferroelectrics with the perovskite
structure have shown that their properties differ strongly from those of
bulk crystals. It was established that substrate-induced strain (epitaxial
strain) has a strong influence on the properties of films. Due to strong
coupling between strain and polarization, this strain changes significantly
the phase transition temperature and can induce unusual polar states in thin
films.~\cite{PhysRevLett.80.1988, PhysRevB.69.212101, Nature.430.758,
RevModPhys.77.1083}

To date, the most experimentally studied ferroelectric superlattice is the
BaTiO$_3$/SrTiO$_3$ (BTO/STO) one.~\cite{JApplPhys.72.2840, ApplPhysLett.65.1784,
ApplPhysLett.65.1970, ApplPhysLett.70.321, ApplPhysLett.72.1394, 
PhysRevB.60.1697, ApplPhysLett.77.3257, JApplPhys.91.2284, JApplPhys.91.2290,
ApplPhysLett.80.3581, ApplPhysLett.82.2118, JPhysCondensMatter.15.L305,
JApplPhys.93.1180, MaterSciEngB.98.6, JApplPhys.94.7923,
SciTechnolAdvMater.5.425, PhysRevB.69.132302,
Ferroelectrics.329.3,Science.313.1614, JApplPhys.100.051613,
ApplPhysLett.89.092905, Ferroelectrics.357.128, ThinSolidFilms.515.6438,
Ferroelectrics.370.57, PhysRevLett.101.197402, PhysRevB.82.224102,
PhysRevB.83.144108, PhysSolidState.53.1062}
Studies of these superlattices from first principles~\cite{ApplPhysLett.82.1586,
PhysRevB.71.100103, IntegratedFerroelectrics.73.3, ApplPhysLett.87.052903,
PhysRevB.72.214121, JApplPhys.100.051613, ApplPhysLett.89.092905,
PhysRevB.76.020102, JApplPhys.103.124106, JPhysD.41.215408, PhysRevB.79.024101,
JApplPhys.105.016104, PhysSolidState.51.2324, PhysSolidState.52.1448}
have established main factors responsible for the formation of their
polar structure. The specific feature of the superlattice is that the strains
induced in it by the lattice mismatch between BaTiO$_3$ and SrTiO$_3$ and by
the substrate result in concurrency of equilibrium polar structures in
neighboring layers, so that the polar structure of the superlattice can be
tetragonal, monoclinic, or orthorhombic, depending on the mechanical boundary
conditions at the interface with the substrate.

Although some properties of BTO/STO superlattices have been already studied,
a number of problems remain unresolved. For instance, first-principles study
of dielectric properties of these superlattices~\cite{ApplPhysLett.87.052903,
PhysRevB.72.214121} have found only $P4mm$ and $Cm$ polar phases, whereas the
$Amm2$ phase, which is characteristic for stretched films of
BaTiO$_3$,~\cite{PhysRevB.69.212101} SrTiO$_3$,~\cite{PhysRevB.71.024102} and
for PbTiO$_3$/PbZrO$_3$ superlattices,~\cite{PhysRevB.69.184101} was not
observed. The piezoelectric properties were calculated only for
PbTiO$_3$/PbZrO$_3$ superlattice;~\cite{PhysRevB.59.12771} for BTO/STO
superlattices these data are absent. Finally, the elastic properties of
ferroelectric superlattices and their behavior at the boundaries between
different polar phases have not been studied at all.

In this work, first-principles density-functional calculations of the phonon
spectra, crystal structure, spontaneous polarization, dielectric, piezoelectric,
and elastic properties for polar phases of (001)-oriented (BTO)$_m$/(STO)$_n$
superlattices (SL $m/n$) with $m = n = {}$1--4 are performed. The influence
of compressive and tensile epitaxial strain on the structure and properties
of polar phases is studied in details for (BTO)$_1$/(STO)$_1$ superlattice.
The stability ranges of tetragonal, monoclinic, and orthorhombic phases are
determined. The critical behavior of static dielectric, piezoelectric, and
elastic tensors at the boundaries between different polar phases are studied.
The ferroelastic type of the phase transitions between the polar phases is
established. In addition, an important question about the thermodynamic
stability of BTO/STO superlattices is considered.

\begin{table*}
\caption{\label{table1}Parameters used for construction of
pseudopotentials.~\cite{PhysSolidState.51.362}  Non-relativistic generation
scheme was used for Sr, Ti, and O atoms, and scalar-relativistic generation
scheme was used for the Ba atom. All parameters are in Hartree atomic units
except for the energy $V_{\rm loc}$, which is in Ry.}
\begin{ruledtabular}
\begin{tabular}{ccccccccccc}
Atom & Configuration      & $r_s$ & $r_p$ & $r_d$ & $q_s$ & $q_p$ & $q_d$ & $r_{\rm min}$ & $r_{\rm max}$ & $V_{\rm loc}$ \\
\hline
Sr   & $4s^24p^64d^05s^0$ & 1.68  & 1.74  & 1.68  & 7.07  & 7.07  & 7.07  & 0.01          & 1.52          &    1.5 \\
Ba   & $5s^25p^65d^06s^0$ & 1.85  & 1.78  & 1.83  & 7.07  & 7.07  & 7.07  & 0.01          & 1.68          &    1.95 \\
Ti   & $3s^23p^63d^04s^0$ & 1.48  & 1.72  & 1.84  & 7.07  & 7.07  & 7.07  & 0.01          & 1.41          &    2.65 \\
O    & $2s^2 2p^4 3d^0$   & 1.40  & 1.55  & 1.40  & 7.07  & 7.57  & 7.07  & ---           & ---           & --- \\
\end{tabular}
\end{ruledtabular}
\end{table*}

The remainder of this paper is organized as follows. In Sec.~\ref{sec2},
we give the details of our calculations. Next, we present the results for the
ground state (Sec.~\ref{sec31}) and the polarization (Sec.~\ref{sec32})
of (BTO)$_n$/(STO)$_n$ superlattices. Dielectric, piezoelectric, and elastic
properties of (BTO)$_1$/(STO)$_1$ superlattice are described in
Secs.~\ref{sec33}, \ref{sec34} and \ref{sec35}, respectively. The
thermodynamic stability of the superlattices is analyzed in
Sec.~\ref{sec36}. The obtained results are discussed in Sec.~\ref{sec4}.

\section{\label{sec2}Calculation details}

The calculations were performed within the first-principles density-functional
theory (DFT) with pseudopotentials and a plane-wave basis set as implemented
in \texttt{ABINIT} software.~\cite{abinit}  The local density approximation
(LDA) for the exchange-correlation functional~\cite{PhysRevB.23.5048} was
used. Optimized separable nonlocal pseudopotentials~\cite{PhysRevB.41.1227}
were constructed using the \texttt{OPIUM} software;~\cite{opium} to improve
the transferability of pseudopotentials, the local potential correction was
added according to Ref.~\onlinecite{PhysRevB.59.12471}. Parameters used for
construction of pseudopotentials are given in Table~\ref{table1}; the results
of testing of these pseudopotentials and other details of calculations can be
found in Ref.~\onlinecite{PhysSolidState.51.362}. The plane-wave cut-off energy
was 30~Ha (816~eV). The integration over the Brillouin zone was performed with
a 8$\times$8$\times$4 Monkhorst--Pack mesh. The relaxations of the atomic
positions and the unit cell parameters were stopped when the Hellmann-Feynman
forces were below $5 \cdot 10^{-6}$~Ha/Bohr (0.25~meV/{\AA}).

The lattice parameters calculated using the pseudopotentials were
$a = 7.3506$~Bohr (3.8898~{\AA}) for bulk SrTiO$_3$ and $a = 7.4923$~Bohr
(3.9648~{\AA}), $c = 7.5732$~Bohr (4.0075~{\AA}) for tetragonal BaTiO$_3$.
Slight underestimation of the lattice parameters (in our case by 0.4--0.7\%
compared to the experimental data) is a known problem of LDA calculations.

The calculations were performed on two structures: supercells of
$1 \times 1 \times 2n$ perovskite unit cells for (001)-oriented
(BTO)$_n$/(STO)$_n$ superlattices ($n = {}$1--4) and two special quasirandom
structures (SQS) for Ba$_{0.5}$Sr$_{0.5}$TiO$_3$ solid solution; the
construction of SQSs is described in Sec.~\ref{sec36}.

Phonon spectra, dielectric, piezoelectric, and elastic properties of the
superlattices were calculated within the DFT perturbation theory. Phonon
contribution to the dielectric constant was calculated from phonon
frequencies and oscillator strengths.~\cite{PhysRevB.63.104305} The Berry
phase method~\cite{PhysRevB.47.1651} was used to calculate the spontaneous
polarization $P_s$.

As the layers of the superlattices are epitaxially grown on (001)-oriented
substrate with a cubic structure, the calculations were performed for
pseudotetragonal unit cells in which two in-plane translation vectors have
the same length $a_0$ and all three translation vectors are perpendicular
to each other. This means that for monoclinic and orthorhombic phases (with
$Cm$ and $Amm2$ space groups) small deviations of the angles between the
translation vectors from 90$^\circ$ (which were typically less than
0.07$^\circ$) were neglected. As was checked, this does not influence
much the results.

\section{Results}

\subsection{\label{sec31}Ground state of epitaxially strained superlattices}

The lattice mismatch between BaTiO$_3$ and SrTiO$_3$ creates tensile biaxial
strain in SrTiO$_3$ layers and compressive biaxial strain in BaTiO$_3$ layers
of free-standing BTO/STO superlattice. If the layers were isolated, these
strains would result in appearance of the in-plane spontaneous polarization
in SrTiO$_3$ ($Amm2$ space group) and in increasing of the out-of-plane
polarization in BaTiO$_3$ layers ($Cm$ or $P4mm$ space
groups).~\cite{PhysRevLett.80.1988, PhysRevB.69.212101, Nature.430.758}
As the polar state with strong local variations of polarization is energetically
unfavorable,~\cite{ApplPhysLett.82.1586} the structure of the polar ground
state of the superlattice requires special consideration. Earlier studies
of BTO/STO~\cite{ApplPhysLett.87.052903,PhysRevB.72.214121} and
PbTiO$_3$/PbZrO$_3$~\cite{PhysRevB.69.184101} superlattices have demonstrated
that both the magnitude and orientation of polarization depend also on the
substrate-induced (epitaxial) strain in superlattices.

The ground state of BTO/STO superlattice was searched as follows. For a set of
in-plane lattice parameters $a_0$, which were varied from 7.35 to 7.50~Bohr,
we first calculated the equilibrium structure of the paraelectric phase with
$P4/mmm$ space group by minimizing the Hellmann-Feynman forces. The phonon
frequencies at the $\Gamma$ point were then calculated for these structures.
It is known that the ground state of any crystal is characterized by positive
values of all optical phonon frequencies at all points of the Brillouin zone.
So, if the structure under consideration exhibited unstable phonons (with
imaginary phonon frequencies), the atomic positions in it were slightly
distorted according to the eigenvector of the most unstable mode, and a new
search for the equilibrium structure was initiated. The phonon frequencies
calculation and the search for equilibrium structure were repeated until the
structure with all positive phonon frequencies was found.

\begin{figure}
\centering
\includegraphics{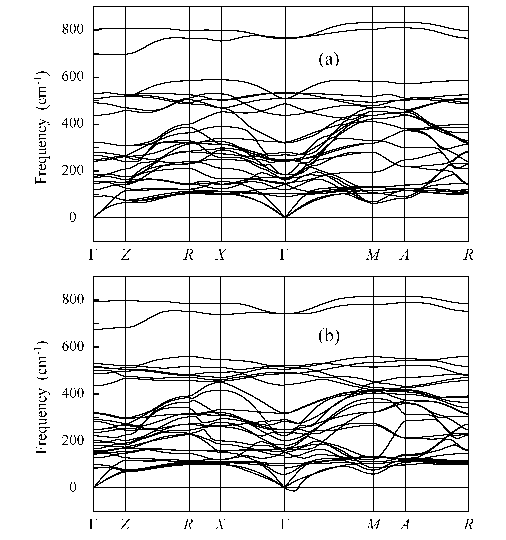}
\caption{\label{fig1}Phonon spectra for (a) the $P4mm$ phase of
(BTO)$_1$/(STO)$_1$ superlattice grown on SrTiO$_3$ substrate
($a_0 = 7.3506$~Bohr) and (b) the $Cm$ phase of the same, free-standing
superlattice ($a_0 = 7.4461$~Bohr).}
\end{figure}

It should be noted that the only unstable mode in the paraelectric $P4/mmm$
phase of (BTO)$_1$/(STO)$_1$ superlattice is the ferroelectric one at the
$\Gamma$ point. The well-known structural instability of SrTiO$_3$ associated
with the $R_{25}$ phonon mode at the boundary of the Brillouin zone disappears
in the superlattice: the frequency of the corresponding phonon at the $M$
point of the folded Brillouin zone (to which the $R$ point transforms when
doubling the $c$ lattice parameter) is 55~cm$^{-1}$ for 1/1 free-standing
superlattice and 61~cm$^{-1}$ for 1/1 superlattice grown on SrTiO$_3$ substrate
(see Fig.~\ref{fig1}).

The phonon spectra calculations show that in 1/1 superlattice grown on
SrTiO$_3$ substrate (compressive epitaxial strain, the in-plane lattice
parameter $a_0$ is equal to that of cubic strontium titanate) the tetragonal
polar phase with $P4mm$ space group is the ground state (Fig.~\ref{fig1}).
For free-standing superlattice, the $P4mm$ structure is unstable and
transforms to monoclinic $Cm$ polar one. Under tensile epitaxial strain
($a_0 = 7.46$~Bohr), the orthorhombic $Amm2$ polar phase is the most stable
one for 1/1 superlattice. This means that the variation of $a_0$ (for example,
by growing the superlattice on different substrates) can be used to control
the polar state of the superlattice.

\begin{figure}
\centering
\includegraphics{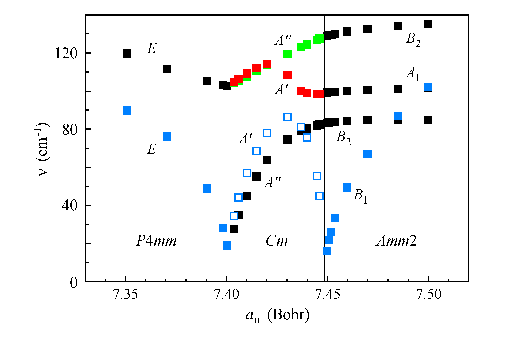}
\caption{\label{fig2}(Color online) Frequencies of four lowest phonon modes
at the $\Gamma$ point for polar phases of (BTO)$_1$/(STO)$_1$ superlattice
as a function of the in-plane lattice parameter $a_0$. The labels near the
points indicate the symmetry of modes. Vertical lines indicate the phase
boundaries.}
\end{figure}

In order to determine accurately the location of the boundaries between $P4mm$
and $Cm$ phases and between $Cm$ and $Amm2$ phases for 1/1 superlattice,
the ground state was calculated for a set of in-plane lattice parameters~$a_0$,
and for each of these structures the phonon frequencies at the $\Gamma$
point were computed. The dependence of four lowest phonon frequencies on
the $a_0$ parameter is plotted in Fig.~\ref{fig2}. It is seen that the
frequency of a doubly degenerate $E$ mode decreases critically when
approaching the boundary between $P4mm$ and $Cm$ phases from the tetragonal
phase. After transition to the monoclinic phase two non-degenerate $A^\prime$
and $A^{\prime\prime}$ soft modes appear in the phonon spectrum; the first
of these modes becomes soft again when approaching the boundary between $Cm$
and $Amm2$ phases. The soft phonon mode in the $Amm2$ phase has the $B_1$
symmetry.

Extrapolation of the squared frequencies of soft ferroelectric modes ($E$
mode in the $P4mm$ phase, $A^{\prime}$ mode in the $Cm$ phase and $B_1$ mode
in the $Amm2$ phase) as a function of $a_0$ to zero gives the in-plane
lattice parameters corresponding to the boundaries between different
polar phases. The $P4mm$--$Cm$ boundary is at $a_0 = 7.4023$~Bohr when
extrapolating from the tetragonal phase and at 7.4001~Bohr when extrapolating
from the monoclinic phase. The $Cm$--$Amm2$ boundary is at $a_0 = 7.4489$~Bohr
when extrapolating from the orthorhombic phase and at 7.4483~Bohr when
extrapolating from the monoclinic phase. Small difference between the
values obtained from extrapolation from two sides of the boundary means
that both phase transitions are close to the second-order ones.
Taking into account that a zero in-plane strain in polar superlattice
corresponds to the in-plane lattice parameter of $a_0 = 7.4462$~Bohr, we get
the values of $-$0.605\% and +0.032\% for the misfit strains corresponding
to the phase boundaries.

Comparison of the energies of different polar phases and calculation of the
phonon frequencies at the $\Gamma$ point for (BTO)$_2$/(STO)$_2$ and
(BTO)$_3$/(STO)$_3$ superlattices shows that the same $P4mm$, $Cm$, and
$Amm2$ phases has the lowest energy, respectively, in superlattices grown
on SrTiO$_3$, free-standing superlattices, and superlattices grown on the
substrate with $a_0 = 7.46$~Bohr.

\subsection{\label{sec32}Spontaneous polarization}

\begin{table*}
\caption{\label{table2}Spontaneous polarization (in C/m$^2$) for BTO/STO
superlattices with different thickness of layers, two SQS-4 structures
used for modeling of disordered Ba$_{0.5}$Sr$_{0.5}$TiO$_3$ solid solution,
and tetragonal barium titanate. The in-plane lattice parameters $a_0$
(in Bohr) for superlattices are also presented.}
\begin{ruledtabular}
\begin{tabular}{cccccccccccc}
Structure       & \multicolumn{2}{c}{SL\,1/1} & \multicolumn{2}{c}{SL\,2/2} & \multicolumn{2}{c}{SL\,3/3} & \multicolumn{2}{c}{SL\,4/4} & SQS-4a & SQS-4b & BaTiO$_3$ \\
\hline
$P_s$ orientation & [$xxz$] & [001]\footnotemark[1] & [$xxz$] & [001]\footnotemark[1] & [$xxz$] & [001]\footnotemark[1] & [$xxz$] & [001]\footnotemark[1] & [111] & [001] & [001] \\
\hline
\rule{0pt}{4mm}%
$P_z$           & 0.061   & 0.277  & 0.113   & 0.293  & 0.146   & 0.302  & 0.157   & 0.307  & 0.130 & 0.206 & 0.259 \\
$P_x = P_y$     & 0.165   & 0      & 0.159   & 0      & 0.150   & 0      & 0.144   & 0      & 0.130 & 0     & 0 \\
$a_0$           & 7.4461  & 7.3506 & 7.4432  & 7.3506 & 7.4403  & 7.3506 & 7.4391  & 7.3506 \\
\end{tabular}
\end{ruledtabular}
\footnotetext[1]{Superlattices grown on SrTiO$_3$ substrate.}
\end{table*}

The calculated spontaneous polarization for free-standing and
substrate-supported superlattices, tetragonal BaTiO$_3$, and disordered
Ba$_{0.5}$Sr$_{0.5}$TiO$_3$ solid solution modeled using SQS-4 structures
(see Sec.~\ref{sec36}) are given in Table~\ref{table2}.

\begin{figure}
\centering
\includegraphics{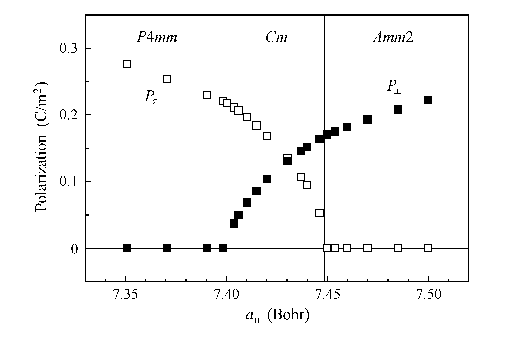}
\caption{\label{fig3}The in-plane ($P_{\perp}$) and out-of-plane ($P_z$)
components of polarization for (BTO)$_1$/(STO)$_1$ superlattice as a
function of the in-plane lattice parameter $a_0$. Vertical lines indicate
the phase boundaries.}
\end{figure}

As was established in Sec.~\ref{sec31}, the tetragonal $P4mm$ phase with
the polarization vector normal to the layers is the ground state for BTO/STO
superlattices grown on SrTiO$_3$ substrate. The calculations shows that in
these superlattices the spontaneous
polarization $P_s$ increases monotonically from 0.277~C/m$^2$ to 0.307~C/m$^2$
as the layer thickness is increased from $n = 1$ to~4 unit cells (see
Table~\ref{table2}). The obtained $P_s$ values agree well with the value of
0.28~C/m$^2$ estimated from the data of Ref.~\onlinecite{ApplPhysLett.82.1586}
for the superlattice with equal thickness of BaTiO$_3$ and SrTiO$_3$ layers;
the $P_s$ value of 0.259~C/m$^2$ for tetragonal BaTiO$_3$ agrees well with the
value of 0.250~C/m$^2$ reported in Ref.~\onlinecite{ApplPhysLett.82.1586}. As
follows from Table~\ref{table2}, for all superlattices the $P_s$ values are
larger than those for Ba$_{0.5}$Sr$_{0.5}$TiO$_3$ solid solution; for
superlattices grown on SrTiO$_3$ substrates they are even larger than $P_s$
of tetragonal BaTiO$_3$. These results agree with
experiment~\cite{JApplPhys.91.2290} and results of previous
calculations.~\cite{ApplPhysLett.82.1586, JApplPhys.105.016104}

For free-standing superlattices, the monoclinic $Cm$ phase is the ground
state; the components of the polarization vector in this phase are also
given in Table~\ref{table2}. It is seen that the polarization is rotated
continuously in the ($\bar{1}10$) plane and the magnitude of polarization
increases with increasing $n$.

For biaxially stretched superlattices, the $Amm2$ phase is the ground state
and the polarization vector is oriented along [110] direction of the reference
tetragonal $P4/mmm$ structure of the paraelectric phase.

The in-plane and out-of-plane components of the polarization for 1/1
superlattice are plotted as a function of the in-plane lattice parameter
$a_0$ in Fig.~\ref{fig3}. Extrapolation of the $P^2_{\perp}$ and
$P^2_z$ dependence on $a_0$ to zero gives the positions
of the $P4mm$--$Cm$ and $Cm$--$Amm2$ phase boundaries. Their values,
$a_0 = 7.4018$~Bohr and $a_0 = 7.4492$~Bohr, are very close to those
obtained in Sec.~\ref{sec31} from the frequencies of soft modes.

\subsection{\label{sec33}Dielectric properties}

\begin{figure}
\centering
\includegraphics{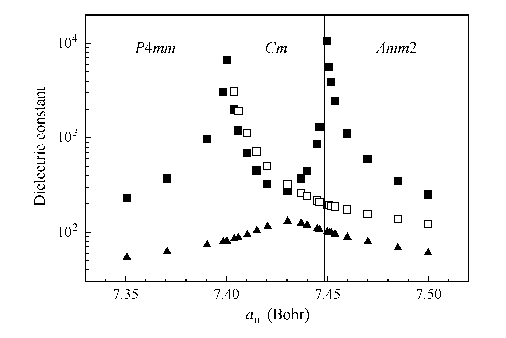}
\caption{\label{fig4}Eigenvalues of the static dielectric constant tensor
$\varepsilon_{ij}$ for (BTO)$_1$/(STO)$_1$ superlattice as a function of
the in-plane lattice parameter $a_0$. Vertical lines indicate the phase
boundaries.}
\end{figure}

The eigenvalues of the static dielectric constant tensor $\varepsilon_{ij}$
($i, j$ = 1, 2, 3) for (BTO)$_1$/(STO)$_1$ superlattice as a function of
the in-plane lattice parameter $a_0$ are shown in Fig.~\ref{fig4}. In the
tetragonal phase, the eigenvectors of the $\varepsilon_{ij}$ tensor coincide
with crystallographic axes and $\varepsilon_{11} = \varepsilon_{22}$. So,
the dielectric properties of this phase are described by two nonzero
independent parameters, $\varepsilon_{11}$ and $\varepsilon_{33}$.

In the monoclinic phase, the polarization vector rotates monotonically in
the ($\bar{1}10$) plane; all three eigenvectors of the $\varepsilon_{ij}$
tensor are different and do not coincide with crystallographic axes of the
reference tetragonal structure. As all nine components of the $\varepsilon_{ij}$
tensor in this coordinate system are nonzero for the $Cm$ phase, the most
compact way to describe the properties of this tensor is to present its
eigenvalues. In the $Cm$ phase, the direction of the eigenvector corresponding
to the smallest eigenvalue is close, but do not coincide with the direction
of the polarization.

In the orthorhombic phase, the eigenvectors of the $\varepsilon_{ij}$ tensor
are oriented along the [110], [$1\bar{1}0$] and [001] directions of the
reference tetragonal structure. The direction of the eigenvector
corresponding to the smallest eigenvalue coincides with the polarization
vector and the direction of eigenvector corresponding to the largest
eigenvalue is [001].

As follows from Fig.~\ref{fig4}, at least one of the eigenvalues of the
$\varepsilon_{ij}$ tensor diverges critically at the $P4mm$--$Cm$ and
$Cm$--$Amm2$ boundaries as the in-plane lattice parameter $a_0$ is changed.
When approaching the $P4mm$--$Cm$ boundary from the tetragonal phase, the
$\varepsilon_{11} = \varepsilon_{22}$ components of this tensor diverge as
the polarization vector $\mathbf{P_s} \parallel {}$[001] becomes less stable
against its rotation in the (${\bar 1}10$) plane. When approaching the
$Cm$--$Amm2$ boundary from the orthorhombic phase, the $\varepsilon_{33}$
value diverges as the polarization vector $\mathbf{P_s} \parallel {}$[110]
becomes less stable against its rotation in the same plane.

\subsection{\label{sec34}Piezoelectric properties}

Due to high sensitivity of both magnitude and orientation of the polarization
vector in superlattices to epitaxial strain, we can expect them to be good
piezoelectrics. It is known that anomalously high piezoelectric moduli found
in some ferroelectrics like PbZr$_{1-x}$Ti$_x$O$_3$ near the morphotropic
phase boundary are due to the ease of strain-induced rotation of polarization
in the intermediate monoclinic phase.~\cite{Nature.403.281,PhysRevLett.84.5423}
A similar situation appears in the monoclinic phase of BTO/STO superlattice.
To our knowledge, the piezoelectric properties of BTO/STO superlattices have
not been studied so far neither experimentally, nor theoretically. The only
superlattice, for which some piezoelectric properties were calculated, is the
PbTiO$_3$/PbZrO$_3$ 1/1 superlattice,~\cite{PhysRevB.68.014112} which was
used to simulate the properties of PbTi$_{0.5}$Zr$_{0.5}$O$_3$ solid solution.

\begin{table}
\caption{\label{table3}Largest piezoelectric moduli for monoclinic phase of
free-standing (BTO)$_1$/(STO)$_1$ superlattice, for tetragonal phase of
the same superlattice grown on SrTiO$_3$ substrate, and for tetragonal barium
titanate.}
\begin{ruledtabular}
\begin{tabular}{cccc}
Structure                  & \multicolumn{2}{c}{SL\,1/1} & BaTiO$_3$ \\
\hline
$P_s$~orientation          & [$xxz$]   & [001]\footnotemark & [001] \\
\hline
\rule{0pt}{4mm}%
$e_{33}$,~C/m$^2$ ($d_{33}$, pC/N) & 31.9 (460)        & 7.1 (49) & 6.3 (42) \\
$e_{15}$,~C/m$^2$ ($d_{15}$, pC/N) & $-$0.09 ($-$19)   & 3.2 (31) & $-$2.9 ($-$24) \\
\end{tabular}
\end{ruledtabular}
\footnotetext{Superlattice grown on SrTiO$_3$ substrate.}
\end{table}

The largest piezoelectric stress moduli $e_{i\nu}$ ($i = 1$, 2, 3; $\nu = 1$--6)
calculated for the $P4mm$ phase of (BTO)$_1$/(STO)$_1$ superlattice grown
on SrTiO$_3$ substrate and for the $Cm$ phase of the same free-standing
superlattice are given in Table~\ref{table3}. It is seen that in tetragonal
phases of the superlattice and BaTiO$_3$ the $e_{33}$ moduli do not differ
much. However in the monoclinic phase, which is the ground state for
free-standing superlattice, the $e_{33}$ value is five times larger. Even
stronger effect can be seen for the $d_{33}$ piezoelectric strain coefficient
($d_{i\nu} = \sum_{\mu=1}^6 e_{i\mu}S_{\mu\nu}$), which is a result of an
1.5-fold increase in the elastic compliance modulus $S_{33}$ in the monoclinic
phase (see Sec.~\ref{sec35}).

\begin{figure}
\centering
\includegraphics{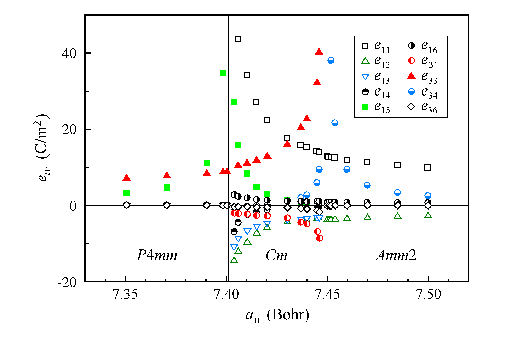}
\caption{\label{fig5}(Color online) Components of the piezoelectric tensor
$e_{i\nu}$ in polar phases of (BTO)$_1$/(STO)$_1$ superlattice as a function
of the in-plane lattice parameter $a_0$. Vertical lines indicate the phase
boundaries.}
\end{figure}

The piezoelectric moduli $e_{i\nu}$ in polar phases of (BTO)$_1$/(STO)$_1$
superlattice as a function of the in-plane lattice parameter are shown in
Fig.~\ref{fig5}. According to the symmetry, in the tetragonal phase the
piezoelectric tensor has three independent and five nonzero components:
$e_{31}=e_{32}$, $e_{33}$, and $e_{15}=e_{24}$. In our superlattice only
two of them have large values: $e_{33}$ and $e_{15}$. When
approaching the $P4mm$--$Cm$ boundary from the tetragonal phase, the
$e_{33}$ value increases monotonically whereas the $e_{15}$ value diverges
critically and reaches the value of 80~C/m$^2$ (not shown).

In the orthorhombic phase (in the coordinate system of the reference tetragonal
structure) the $e_{11}=e_{22}$, $e_{12}=e_{21}$, $e_{13}=e_{23}$,
$e_{34}=e_{35}$, and $e_{16}=e_{26}$ moduli are nonzero, and the total number
of independent parameters is~5. Among them the $e_{11}$, $e_{12}$, $e_{34}$,
and $e_{16}$ moduli have the largest values (see Fig.~\ref{fig5}). The only
modulus that behaves critically at the boundary between $Cm$ and $Amm2$ phases
is the $e_{34}$ one, with a maximum value reaching 192~C/m$^2$ (not shown).

The most complex behavior of piezoelectric moduli is observed in the monoclinic
phase because the change of the in-plane lattice parameter $a_0$ results in
monotonic rotation of the polarization vector in the (${\bar 1}10$) plane. In
the coordinate system of the reference tetragonal structure, all 18~components
of the $e_{i\nu}$ tensor are nonzero (the total number of independent
parameters is 10). As follows from Fig.~\ref{fig5}, in the monoclinic
phase the $e_{11}=e_{22}$, $e_{15}=e_{24}$, $e_{12}=e_{21}$, $e_{13}=e_{23}$,
$e_{14}=e_{25}$, and $e_{16}=e_{26}$ moduli diverge critically at the
boundary between $Cm$ and $P4mm$ phases, and $e_{33}$, $e_{34}=e_{35}$,
$e_{31}=e_{32}$, and $e_{36}$ moduli diverge critically at the boundary
between $Cm$ and $Amm2$ phases. It should be noted that when crossing the
boundaries, small additional jumps (about $\sim$10\%) are observed in the
$e_{33}$ modulus (at the $Cm$--$P4mm$ boundary) and in $e_{11}$ and $e_{12}$
moduli (at the $Cm$--$Amm2$ boundary).

\subsection{\label{sec35}Elastic properties}

It is known that ferroelectric phase transitions between two polar phases
are often the \emph{improper ferroelastic} ones, which means that they are
accompanied by appearance of soft \emph{acoustic} modes and spontaneous
strain, but the strain is not the primary order parameter. This occurs when
the strain tensor and the polar vector transform according to the same
irreducible representation of the high-symmetry phase.~\cite{Blinc}
As such phase transitions occur in BTO/STO superlattices when changing the
in-plane lattice parameter $a_0$, it was interesting to study the influence
of these phase transitions on the elastic properties of superlattices,
especially taking into account that these properties of superlattices have
not been studied so far.

In the tetragonal $P4mm$ phase, the elastic compliance tensor $S_{\mu\nu}$
($\mu,\nu = 1$--6) has six independent and nine nonzero components. In the
orthorhombic $Amm2$ phase (in the coordinate system of the reference tetragonal
structure) the tensor has 9~independent and 13~nonzero components, and in
the monoclinic $Cm$ phase it has 13~independent and 21~nonzero components.

\begin{figure}
\medskip
\centering
\includegraphics{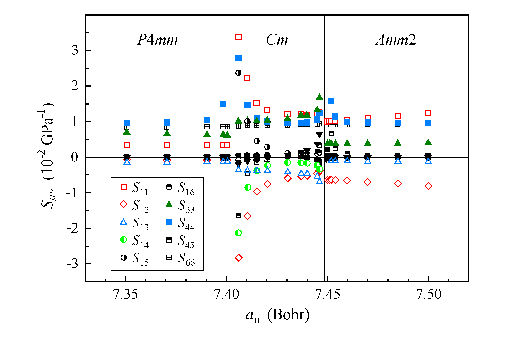}
\caption{\label{fig6}(Color online) Components of the elastic compliance
tensor $S_{\mu\nu}$ for polar phases of (BTO)$_1$/(STO)$_1$ superlattice as
a function of the in-plane lattice parameter $a_0$. Vertical lines indicate
the phase boundaries.}
\end{figure}

The components of the elastic compliance tensor $S_{\mu\nu}$ for polar
phases of (BTO)$_1$/(STO)$_1$ superlattice are plotted as a function of the
in-plane lattice parameter $a_0$ in Fig.~\ref{fig6}. It is seen that at
the boundary between $P4mm$ and $Cm$ phases the components of $S_{\mu\nu}$
tensor exhibit a step-like change ($S_{13}=S_{23}$, $S_{33}$, $S_{66}$,
$S_{16}=S_{26}$, $S_{36}$, $S_{34}=S_{35}$, $S_{46}=S_{56}$ moduli),
a critical divergence from the monoclinic side ($S_{11}=S_{22}$, $S_{12}$,
$S_{15}=S_{24}$, $S_{14}=S_{25}$, $S_{45}$ moduli), or critical divergences
from both sides of the boundary ($S_{44}=S_{55}$ modulus).
In the monoclinic phase, the $S_{14}$, $S_{15}$, $S_{16}$, $S_{34}$,
$S_{36}$, $S_{45}$, and $S_{46}$ moduli become nonzero. In the orthorhombic
phase, the $S_{14}$, $S_{15}$, $S_{34}$, and $S_{46}$ moduli vanish again
whereas the other moduli remain nonzero.
At the boundary between $Cm$ and $Amm2$ phases the anomalies in elastic
moduli are smaller: the $S_{11}$, $S_{12}$, and $S_{66}$ moduli exhibit
step-like changes, the $S_{13}$, $S_{14}$, $S_{15}$, $S_{16}$, $S_{33}$,
$S_{34}$, $S_{36}$, and $S_{46}$ moduli exhibit weak divergence from the
monoclinic side, and the $S_{44}$ and $S_{45}$ moduli exhibit weak divergence
from both sides of the boundary.

The critical divergence of the $S_{44}$ modulus at both $P4mm$--$Cm$ and
$Cm$--$Amm2$ boundaries indicates that the phase transitions induced in
the superlattice by the increase of the in-plane lattice parameter are
indeed the ferroelastic ones.

\subsection{\label{sec36}Thermodynamic stability}

Thermodynamic stability of ferroelectric superlattices is very important
for their possible applications. Thermodynamic stability of BTO/STO
superlattice is determined by the mixing enthalpy of the superlattice and
its relationship with the mixing enthalpy of the disordered
Ba$_{0.5}$Sr$_{0.5}$TiO$_3$ solid solution. The most complex part of the
first-principles calculation of thermodynamic stability is the calculation
of the mixing enthalpy for a solid solution because its simulation using
supercells with a large number of randomly distributed atoms makes it
extremely time-consuming.

\begin{table*}
\caption{\label{table4}Translation vectors, superlattice axis, stacking
sequence of atomic planes, and correlation functions $\overline\Pi_{2,m}$
for SQS-4 structures used for modeling of disordered
Ba$_{0.5}$Sr$_{0.5}$TiO$_3$ solid solution.}
\begin{ruledtabular}
\begin{tabular}{ccccccc}
Structure & Translation vectors & Axis and stacking sequence & $\overline\Pi_{2,1}$ & $\overline\Pi_{2,2}$ & $\overline\Pi_{2,3}$ & $\overline\Pi_{2,4}$ \\
\hline
\rule{0pt}{4.2mm}%
SQS-4a & [$2\bar{1}1$],\,[$1\bar{1}2$],\,[$1\bar{2}1$] & [$1\bar{1}1$] \emph{AABB} & 0 & 0 & 0 & $-$1 \\
SQS-4b & [210],\,[$\bar{2}$10],\,[001]                 & [$\bar{1}20$] \emph{AABB} & 0 & $-$1/3 & 0 & 1/3 \\
\end{tabular}
\end{ruledtabular}
\end{table*}

A conceptually new approach to this problem was proposed by Zunger
\emph{et al}.~\cite{PhysRevLett.65.353} In this approach, a disordered solid
solution $A_xB_{1-x}$ is modeled using a special quasirandom structure
(SQS)---a short-period superstructure, whose statistical properties
(numbers of $N_{AA}$, $N_{BB}$, and $N_{AB}$ atomic pairs in few nearest
shells) are as close as possible to those of ideal disordered solid solution
(at the same time the sites of the superstructure are deterministically filled
with \emph{A} and \emph{B} atoms). This method has been widely used to study
the electronic structure and physical properties of semiconductor solid
solutions and the ordering phenomena in metal alloys. To study the
properties of ferroelectric solid solutions this approach was used quite
rare.~\cite{Ferroelectrics.270.173, PhysRevB.70.134110}

\begin{figure}
\centering
\includegraphics{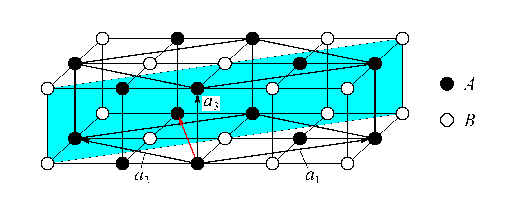}
\caption{\label{fig7}(Color online) The unit cell of SQS-4b superstructure
contains four primitive cells of the perovskite structure and is constructed
using $\bf{a}_1$, $\bf{a}_2$ and $\bf{a}_3$ translation vectors. The atoms of
one of two types, $A$ or $B$, occupy the sites of one sublattice of the
perovskite structure lying on the planes (shown by blue), which are
perpendicular to the superlattice axis (shown by red arrow). The stacking
sequence of the planes along the superlattice axis is $AABB$.}
\end{figure}

The structure of the disordered Ba$_{0.5}$Sr$_{0.5}$TiO$_3$ solid solution
was modeled using two special quasirandom structures SQS-4 constructed with
the \texttt{gensqs} program from \texttt{ATAT}
toolkit.~\cite{JPhaseEquil.23.348} One of these structures is sketched in
Fig.~\ref{fig7}. The translation vectors, superlattice axis, and stacking
sequence of the planes filled with the same atoms, Ba or Sr (denoted by $A$
and $B$), are given in Table~\ref{table4}. The pair correlation functions
$\overline\Pi_{2,m}$ ($m$ is the shell number), which describe the deviation
of statistical properties of these SQSs from those of an ideal solid
solution, are also given in this table. For $x = 0.5$ $\overline\Pi_{2,m}$
is simply ($2N_{AA} / N_m - 1$), where $N_m$ is a number of neighbors in
the $m$th shell. As follows from this table, for the SQS-4a structure strong
deviation from an ideal solid solution appears only in the forth shell; for
the SQS-4b structure deviations appear in the second and fourth shells, but
are smaller. The mixing enthalpy $\Delta H$ for all studied structures $X$
(superlattices with different periods and SQS structures) was calculated
using the formula
    $$\Delta H = E_{\rm tot}(X) - [E_{\rm tot}(\rm{BaTiO_3}) + E_{\rm tot}(\rm{SrTiO_3})]/2$$
from the values of the total energy $E_{\rm tot}$ (per five-atom formula
unit) for free-standing fully relaxed paraelectric $Pm3m$ and $P4/mmm$ phases.
The obtained values of $\Delta H$ for these structures are given in
Table~\ref{table5}.

\begin{table}
\caption{\label{table5}The mixing enthalpy (in meV) for five
(BTO)$_n$/(STO)$_n$ superlattices with different periods and two SQS-4
structures used for modeling of disordered Ba$_{0.5}$Sr$_{0.5}$TiO$_3$
solid solution.}
\begin{center}
\begin{ruledtabular}
\begin{tabular}{ccccccc}
 SL\,1/1 & SL\,2/2 & SL\,3/3 & SL\,4/4 & SL\,5/5 & SQS-4a & SQS-4b \\
\hline
 2.9     & 8.9     & 11.4    & 12.6    & 13.4    & 16.8  & 11.0 \\
\end{tabular}
\end{ruledtabular}
\end{center}
\end{table}

An unexpected result of our calculation is the fact that $\Delta H$ values
for two shortest-period superlattices (1/1 and 2/2) appeared smaller than
$\Delta H$ values for both realizations of disordered solid solution. This
means that a tendency to \emph{short-range ordering} of components exists
in the BaTiO$_3$--SrTiO$_3$ system. Low values of $\Delta H$ for these
superlattices ($<9$~meV) indicate that short-period BTO/STO superlattices
are thermodynamically stable at 300~K.

The tendency to short-range ordering found in (001)-oriented BTO/STO
superlattices can be explained by a general tendency of the $A$ cations to
order in a layered manner in perovskites, in contrast to the $B$ cations,
which prefer a rock-salt ordering.~\cite{JMaterChem.20.5785}
One can add that a similar effect was observed in our studies of
(001)-oriented (PbTiO$_3$)$_n$/(SrTiO$_3$)$_n$ superlattices, where
\emph{negative} values of $\Delta H$ for $n = {}$1--3 and positive values
for larger $n$ were observed.

Our values of the mixing enthalpy for BTO/STO superlattices are much smaller
than $\Delta H$ value obtained in Ref.~\onlinecite{PhysRevB.71.014111}
(42~meV per formula unit). Analysis of the calculation technique used
in Ref.~\onlinecite{PhysRevB.71.014111} shows that atomic positions in the
superstructures were not relaxed and the superstructures were assumed to
be cubic when calculating the energies of different atomic configurations.
So, the calculated mixing enthalpy in this paper includes a large energy of
excess strain.

\section{\label{sec4}Discussion}

Our results on the influence of epitaxial strain on the ground state of
BTO/STO superlattice agree only partially with the results obtained
in Refs.~\onlinecite{ApplPhysLett.87.052903, PhysRevB.71.100103,
PhysRevB.72.214121, JApplPhys.100.051613}. The results coincide in that:
1)~the ground state for free-standing (BTO)$_1$/(STO)$_1$ superlattice is the
monoclinic $Cm$ phase, 2)~under the compressive strain, the superlattice
undergoes the phase transition from $Cm$ to $P4mm$ phase, and 3)~the
dielectric constant diverges at the $P4mm$--$Cm$ phase boundary. At the same
time, in contrast to the results of Refs.~\onlinecite{ApplPhysLett.87.052903,
PhysRevB.72.214121}, we succeeded to observe the phase transition to the $Amm2$
orthorhombic phase under the tensile strain. (In
Refs.~\onlinecite{ApplPhysLett.87.052903, PhysRevB.72.214121} only the
rotation of the polarization vector towards the [110] direction was observed
under the tensile strain.) We consider our results to be more reliable because
they agree with the results obtained for strained
BaTiO$_3$~\cite{PhysRevB.69.212101} and SrTiO$_3$~\cite{PhysRevB.71.024102}
thin films, the results of recent atomistic calculations of the
strain--temperature phase diagram for (BTO)$_2$/(STO)$_2$
superlattice,~\cite{PhysRevB.76.020102} and with results obtained for another
superlattice, PbTiO$_3$/PbZrO$_3$.~\cite{PhysRevB.69.184101} In all these
systems the same phase sequence, $P4mm$--$Cm$--$Amm2$, was observed as the
in-plane lattice parameter was increased.

The increase in the polarization $P_s$ in (BTO)$_n$/(STO)$_n$ superlattices
grown on SrTiO$_3$ substrates with increasing the layer thickness $n$
(Table~\ref{table2}) agrees with the results of
Ref.~\onlinecite{JApplPhys.105.016104} in which the explanation of this
phenomenon was proposed. In free-standing
superlattices, the $P_z$ component of polarization also increased with
increasing $n$, but the $P_x$ and $P_y$ components decreased with increasing
$n$, in contrast to the results observed for 3/3, 4/4 and 5/5 superlattices
with fixed in-plane lattice parameter equal to 1.01 times the lattice parameter
of SrTiO$_3$.~\cite{PhysRevB.71.100103} We attribute these changes to the
decrease of the in-plane lattice parameter $a_0$ for free-standing
superlattices with increasing $n$ (see Table~\ref{table2}).

Unfortunately, only a few data points presented in Ref.~\onlinecite{PhysRevB.72.214121}
for the dielectric constant at the $P4mm$--$Cm$ boundary for (BTO)$_1$/(STO)$_1$
superlattice did not enabled us to make detailed comparison between our results.
However, the comparison of our dielectric data with those calculated for
(PbTiO$_3$)$_1$/(PbZrO$_3$)$_1$ superlattice~\cite{PhysRevB.69.184101} shows
that in the monoclinic phase all eigenvalues of the $\varepsilon_{ij}$ tensor
for BTO/STO superlattice are higher, and so this superlattice may be more
promising for different applications.

To obtain large piezoelectric moduli necessary for technological applications,
the epitaxial strain in the BTO/STO superlattice should be tuned to a value
at which $e_{i\nu}$ and $S_{\mu\nu}$ properties of the superlattice diverge.
As was shown in Secs.~\ref{sec34} and \ref{sec35}, this appears at the
phase boundaries. The calculations shows that in the vicinity of the
$P4mm$--$Cm$ boundary the $d_{11}$ piezoelectric coefficient reaches a maximum
value of 2300~pC/N in the monoclinic phase and the $d_{15}$ coefficient reaches
a value of 6200~pC/N in the tetragonal phase. In the vicinity of the
$Cm$--$Amm2$ boundary, the $d_{33}$ piezoelectric coefficient reaches a value
of 920~pC/N in the monoclinic phase and the $d_{34}$ coefficient reaches a
value of 10500~pC/N in the orthorhombic phase. For comparison, the maximum
piezoelectric coefficient obtained experimentally on single crystals of the
Pb(Zn$_{1/3}$Nb$_{2/3}$)O$_3$--PbTiO$_3$ system was 2500~pC/N.~\cite{JApplPhys.82.1804}
Anomalous increase in the calculated $d_{15}$ and $d_{33}$ coefficients up
to $\sim$8500~pC/N was predicted for hydrostatically stressed PbTiO$_3$ in
the vicinity of the $P4mm$--$Cm$ phase boundary.~\cite{PhysRevLett.95.037601}

\begin{figure}
\centering
\includegraphics{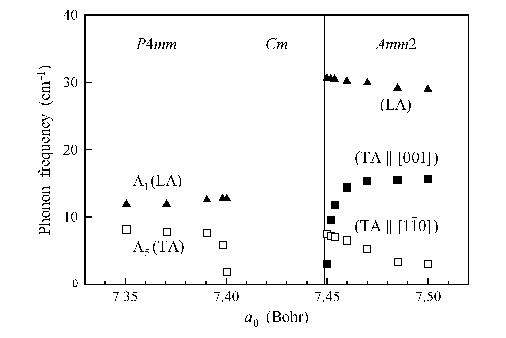}
\caption{\label{fig8}Frequencies of acoustic modes in the vicinity of the
$\Gamma$ point in polar phases of (BTO)$_1$/(STO)$_1$ superlattice as a
function of the in-plane lattice parameter $a_0$. Vertical lines indicate
the phase boundaries.}
\end{figure}

Consider now the elastic properties of BTO/STO superlattice and the results
indicating the appearance of improper ferroelastic phase transitions. According
to Ref.~\onlinecite{Blinc}, in crystals with $P4mm$ space group the phase
transition $4mm$\,$\to$\,$m$ should be of the ferroelastic type. The spontaneous
strain at this phase transition is characterized by one or both nonzero $u_4$
and $u_5$ components of the strain tensor, and a soft transverse acoustic (TA)
mode with a wave vector parallel to the polar axis should appear in the
phonon spectrum in the vicinity of the $\Gamma$ point. Direct calculations
of frequencies of acoustic modes in the tetragonal phase at the point with
a reduced wave vector $\bf{q} = {}$(0,\,0,\,0.05) confirmed this (see
Fig.~\ref{fig8}): when approaching the $P4mm$--$Cm$ boundary, the frequency of
a doubly degenerate TA phonon with the $\Lambda_5$ symmetry decreased by
5~times. The softening of this mode is directly related to the divergence of
the $S_{44}=S_{55}$ components of the elastic compliance tensor.

At the boundary between $Cm$ and $Amm2$ phases, there should be another
ferroelastic phase transition. According to Ref.~\onlinecite{Blinc}, the
transition $mm2$\,$\to$\,$m$ is accompanied by spontaneous strain with
one nonzero of three $u_4$, $u_5$, $u_6$ components. In our coordinate
system with unusual for orthorhombic phase orientation of the polar axis
(along the [110] direction) there should be a softening of TA phonon
with a wave vector oriented along this axis. This was confirmed by direct
calculations of frequencies of acoustic modes at the point in the
Brillouin zone with a reduced wave vector $\bf{q} = {}$(0.035,\,0.035,\,0)
(see Fig.~\ref{fig8}). In contrast to the tetragonal phase in which
the soft acoustic mode is doubly degenerate, in the orthorhombic phase
the only phonon mode that softens at the phase boundary is the TA mode
polarized along the [001] axis. In our opinion, this difference is the
reason why the anomalies in the elastic properties at the ferroelastic
$Cm$--$Amm2$ phase transition are much weaker than those at the ferroelastic
$P4mm$--$Cm$ phase transition. Unusual orientation of the polar axis
in our coordinate system results in coupling of some components of
the elastic compliance tensor: for example, the $S_{44}$ and $S_{45}$
moduli, which diverge in the orthorhombic phase, satisfy the relation
$S_{44} - S_{45} \approx {}$const in this phase.

Negative value of the acoustic phonon frequency in the $Cm$ phase, which
is seen in a narrow wave vector region in Fig.~\ref{fig1}, is an artifact
of calculations. Computation of the phonon dispersion curves in the vicinity
of the $\Gamma$ point revealed three acoustic branches $\omega(q)$, whose
frequencies increased monotonically with increasing $q$, but gave negative
values of $\omega(0)$ (about 8 cm$^{-1}$) in the limit $q \to 0$ because of
numerical errors. After application of the acoustic sum rule,
$\omega(0)$ restored its zero value, but the derivative $d\omega / dq$ near
$q = 0$ for the softest mode became negative. So, there is no contradiction
between the phonon spectra and the positive definiteness of the elastic
moduli matrix $C_{ij}$ calculated in Sec.~\ref{sec35}.

\section{Conclusions}

In this work, the properties of (001)-oriented (BaTiO$_3$)$_m$/(SrTiO$_3$)$_n$
superlattices with $m = n = {}$1--4 were calculated using the first-principles
density-functional theory. For free-standing superlattices, the ground state
is the monoclinic $Cm$ polar phase. Under the in-plane compressive epitaxial
strain, it transforms to tetragonal polar $P4mm$ phase, and under in-plane
tensile strain, it transforms to orthorhombic $Amm2$ polar phase. All components
of the static dielectric tensor ($\varepsilon_{ij}$), the piezoelectric
tensor ($e_{i\nu}$), and the elastic compliance tensor ($S_{\mu\nu}$) were
calculated as a function of the in-plane lattice parameter for 1/1 superlattice.
The critical behavior of some components of $\varepsilon_{ij}$, $e_{i\nu}$,
and $S_{\mu\nu}$ tensors at the boundaries between different polar phases was
observed. It was shown that the phase transitions between different polar
phases are of the improper ferroelastic type. The possibility of obtaining
ultrahigh piezoelectric moduli using fine tuning of the epitaxial strain in
superlattices was demonstrated. The comparison of the mixing enthalpy calculated
for superlattices and disordered Ba$_{0.5}$Sr$_{0.5}$TiO$_3$ solid solution
modeled using two special quasirandom structures SQS-4 revealed a tendency
of the BaTiO$_3$--SrTiO$_3$ system to short-range ordering of cations and
showed that short-period superlattices are thermodynamically quite stable.

\begin{acknowledgments}
This work was supported by the RFBR Grant No.~08-02-01436.
\end{acknowledgments}


\providecommand{\BIBYu}{Yu}

\end{document}